\documentclass[aps,preprint]{revtex4}%
\usepackage{amsfonts}
\usepackage{amsmath}
\usepackage{amssymb}
\usepackage{graphicx}%
\providecommand{\U}[1]{\protect\rule{.1in}{.1in}}

\begin{document}
\preprint{ }
\title[ ]{Quarkonium Masses in a hot QCD\ Medium Using Conformable Fractional of the
Nikiforov-Uvarov Method}
\author{M. Abu-Shady}
\affiliation{{\small Department of Applied Mathematics, Faculty of Science, Menoufia
University, Shbien El-Kom, Egypt}}
\author{}
\affiliation{}
\author{}
\affiliation{}

\begin{abstract}
By using conformable fractional of the Nikiforov-Uvarov (CF-NU) method, the
radial Schr\"{o}dinger equation is analytically solved. The energy eigenvalues
and corresponding functions are obtained, in which the dependent temperature
potential is employed. The effect of fraction-order parameter is studied on
heavy-quarkonium masses such as charmonium and bottomonium in a hot QCD medium
in the 3D and the higher dimensional space. A comparison is studied with
recent works. We conclude that the fractional-order plays an important role in
a hot QCD medium in the 3D and higher-dimensional space.


\end{abstract}
\volumeyear{ }
\volumenumber{ }
\issuenumber{ }
\eid{ }
\startpage{1}
\endpage{102}
\maketitle

\section{Introduction}

The color screening of the static chromo-electric fields is one of the most
remarkable features of the quark-gluon plasma formation. The Debye screening
in quantum chromodynamic theory plasma has been studied as a probe of
deconfinement in a hot medium which shows a reduction in the interaction
between heavy quarks and antiquarks due to color screening leading to a
suppression in $J/\Psi\ $yields [1, 2]. Thereby the quarkonium in the hot
medium is a good tool to examine the confined/deconfined state of matter. The
dissociation of heavy quarkonium states in a hot QCD medium is studied by
investigating the medium modifications to a heavy-quark potential [3].

The development of the Schr\"{o}dinger equation (SE) plays a major role at
finite temperature. Matsui and Satz $\left[  4\right]  $ have investigated the
formation of a hot quark-gluon plasma by calculating the $J/\Psi$ radius of
charmonium. In Ref. $\left[  5\right]  $, the SE is solved by using the
Funke-Hecke theorem to describe the electron and proton media. Wong $\left[
6\right]  $ has studied the binding energies and wave functions of heavy
quarkonia in quark-gluon plasma by using a temperature dependent potential
inferred from lattice gauge calculations. Thus, the study shows that the model
with the new Q-\={Q} potential gives dissociation temperatures that agree with
the spectral function analyses. The SE in a screened Coulomb heavy-quark
potential is solved to\ study the temperature dependence of the
heavy-quarkonium interaction based on the Bhanot--Peskin leading order
perturbative QCD analysis in which the 1S charmonium thermal width is
determined and compared to recent lattice QCD results $\left[  7\right]  $.
Wong $\left[  8\right]  $ has investigated the Q-\={Q} potential by using the
thermodynamic quantities to give spontaneous dissociation temperatures for
quarkonium and has also found the quark drip lines which separate the region
of bound color-singlet Q\={Q} states from the unbound region. In Ref. $\left[
9\right]  ,$ the authors numerically solved the SE at finite temperature by
employing temperature-dependent effective potential given by a linear
combination of color singlet and internal energies. By using thermodynamic
laws, the SE equation is derived at finite temperature $\left[  10\right]  $.

Some authors $\left[  11-17\right]  $ focus to extend the SE to the
higher-dimensional space which gives more detail about the systems under
study. Moreover, the energy eigenvalues and wave functions are obtained in the
higher-dimensional space and their applications on quarkonium properties are
studied in the vacuum, hot and dense media.

Recently, the fractional calculus has attracted attention in the different
fields of physics which the nonlinear, complex effects are included such as in
Refs. $\left[  18-21\right]  $. In high energy physics, the description of
heavy-quarkonium energy spectra and complex phenomena of the standard model as
in Ref. $\left[  22\right]  $, in which, the author used the conformable
fractional derivative to express the fractional radial SE in the N-dimensional
space for the extended Cornall potential by using generalized extended
Nikiforov-Uvarov (ENU) method to the fractional domain. In Ref. $\left[
23\right]  ,$ the fractional form of the NU method is applicable\ in order to
solve fractional radial SE with its applications on variety of potentials such
as the oscillator potential, Woods-Saxon potential, and Hulthen potential. In
Ref. $\left[  24\right]  $, the Caputo fractional derivative is applied on a
fractional Schr\"{o}dinger wave equation with using quantization of the
classical nonrelativistic Hamiltonian. The free particle solutions are
obtained, which are confined to a certain region of space.

Thus, The fractional calculus focuses on vacuum space without considering the
effects of medium. Thus, the present study is to study the effect of
fractional-order on quarkonium masses in a hot medium which are not considered
in recent works such as in Refs. $\left[  18-24\right]  $, The conformable
fractional of the Nikiforov-Uvarov (CF-NU) method is applied to obtain the
analytic solutions of the $N$-dimensional radial SE.

The paper is organized as follows: In Sec. 2, the CF method is briefly
explained. In Sec. 3, The energy eigenvalue and wave function are calculated
in the $N$-dimensional space using CF-NU method. In Sec. 4, the results are
discussed. In Sec. 5, the summary and conclusion are presented.

\section{2. Theoretical Tools}

\subsection{2.1. Conformable fractional derivative}

Fractional derivative plays an important role in applied science.
Riemann--Liouville and Riesz and Caputo give a good formula that allows to
apply boundary and initial conditions as in Ref. $\left[  19\right]  $%

\begin{equation}
D_{t}^{\alpha}f\left(  t\right)  =\int_{t_{0}}^{t}K_{\alpha}\left(
t-s\right)  f^{\left(  n\right)  }\left(  s\right)  ds,\ \ \ \ t>t_{0}\tag{1}%
\end{equation}
with%
\begin{equation}
K_{\alpha}\left(  t-s\right)  =\frac{\left(  t-s\right)  ^{n-\alpha-1}}%
{\Gamma\left(  n-\alpha\right)  },\tag{2}%
\end{equation}
where $f^{\left(  n\right)  }$ is the $n$ th derivative of the function
$f\left(  t\right)  $, and $K_{\alpha}\left(  t-s\right)  $ is the kernel,
which is fixed for a given real number $\alpha$. The kernel $K_{\alpha}\left(
t-s\right)  $ has singularity at $t=s$. Caputo and Fabrizio $\left[
25\right]  $ suggested a new formula of the fractional derivative with smooth
exponential kernel of the form to avoid the difficulties that found in Eq.
$\left(  1\right)  $%

\begin{equation}
D_{t}^{\alpha}f\left(  t\right)  =\frac{M\left(  \alpha\right)  }{1-\alpha
}\int_{t_{0}}^{t}\exp\left(  \frac{\alpha\left(  t-s\right)  }{1-\alpha
}\right)  \dot{y}\left(  s\right)  ds, \tag{3}%
\end{equation}
where $M\left(  \alpha\right)  $ is a normalization function with
$M(0)=M(1)=1$. A new formula of fractional derivative, called conformable
fractional derivative (CFD) is proposed by Khalil et al. $\left[  26\right]  $%
\begin{equation}
D_{t}^{\alpha}f\left(  t\right)  =\lim_{\epsilon\rightarrow0}\frac{f\left(
t+\epsilon t^{1-\alpha}\right)  -f\left(  t\right)  }{\epsilon}\ \ \ t>0
\tag{4}%
\end{equation}

\begin{equation}
f\left(  0\right)  =\lim_{\epsilon\rightarrow0}f\left(  t\right)  \tag{5}%
\end{equation}
with $0<\alpha\leq1$. This new definition is simple and provides a natural
extension of differentiation with integer order $n\in Z$ \ to fractional order
$\alpha\in C$. Moreover, the CFD operator is linear and satisfies the
interesting properties that traditional fractional derivatives do not, such as
the formula of the derivative of the product or quotient of two functions and
the chain rule $\left[  23\right]  .$ The concept of CFD has successfully
applied to the CF-NU method to obtain the eigenvalues and eigenfunctions of SE
as in Refs. $\left[  22,23\right]  .$

\section{Conformable Fractional NU method}

In this section, the CF-NU method is briefly given to solve the conformable
fractional of differential equation which takes the following form \{see Ref.
[23], for details\}
\begin{equation}
D^{\alpha}\left[  D^{\alpha}\Psi(s)\right]  +\frac{\bar{\tau}(s)}{\sigma
(s)}D^{\alpha}\Psi(s)+\frac{\tilde{\sigma}(s)}{\sigma^{2}(s)}\Psi(s)=0,
\tag{6}%
\end{equation}
where $\sigma(s)$ and $\tilde{\sigma}(s)$ are polynomials of maximum second
degree and $\bar{\tau}(s)$ is a polynomial of maximum first degree with an
appropriate $s=s(r)$ coordinate transformation.%
\begin{equation}
D^{\alpha}\Psi(s)=s^{1-\alpha}\Psi^{\prime}(s), \tag{7}%
\end{equation}

\begin{equation}
D^{\alpha}\left[  D^{\alpha}\Psi(s)\right]  =\left(  1-\alpha\right)
s^{1-2\alpha}\Psi^{\prime}(s)+s^{2-2\alpha}\Psi^{"}(s). \tag{8}%
\end{equation}
Substituting by Eqs. (7) and (8) into (6), we obtain%

\begin{equation}
\Psi^{\prime\prime}(s)+\frac{\bar{\tau}_{f}(s)}{\sigma_{f}(s)}\Psi^{\prime
}(s)+\frac{\tilde{\sigma}_{f}(s)}{\sigma_{f}^{2}(s)}\Psi(s)=0,\tag{9}%
\end{equation}
To find particular solution of Eq. (9) by separation of variables, if one
deals with the transformation%
\begin{equation}
\Psi(s)=\Phi(s)\chi(s),\tag{10}%
\end{equation}
it reduces to an equation of hypergeometric type as follows%
\begin{equation}
\sigma_{f}(s)\chi^{\prime\prime}(s)+\tau_{f}(s)\chi^{\prime}(s)+\lambda
\chi(s)=0,\tag{11}%
\end{equation}
where%
\begin{equation}
\sigma_{f}(s)=\pi_{f}(s)\frac{\Phi(s)}{\Phi^{\prime}(s)},\tag{12}%
\end{equation}%
\begin{equation}
\tau_{f}(s)=\bar{\tau}_{f}(s)+2\pi_{f}(s);~\ \ \tau_{f}^{\prime}(s)<0,\tag{13}%
\end{equation}
and
\begin{equation}
\lambda=\lambda_{n}=-n\tau_{f}^{\prime}(s)-\frac{n(n-1)}{2}\sigma_{f}%
^{\prime\prime}(s),n=0,1,2,...\tag{14}%
\end{equation}
$\chi(s)=\chi_{n}(s)$ which is a polynomial of $n$ degree which satisfies the
hypergeometric equation, taking the following form%
\begin{equation}
\chi_{n}(s)=\frac{B_{n}}{\rho_{n}}\frac{d^{n}}{ds^{n}}(\sigma_{f}%
^{\prime\prime}(s)\rho(s)),\tag{15}%
\end{equation}
where $B_{n}$ is a normalization constant and $\rho(s)$ is a weight function
which satisfies the following equation
\begin{equation}
\frac{d}{ds}\omega(s)=\frac{\tau(s)}{\sigma_{f}(s)}\omega(s);\ \ \ \omega
(s)=\sigma_{f}(s)\rho(s),\tag{16}%
\end{equation}%
\begin{equation}
\pi_{f}(s)=\frac{\sigma_{f}^{\prime}(s)-\bar{\tau}_{f}(s)}{2}\pm\sqrt
{(\frac{\sigma_{f}^{\prime}(s)-\bar{\tau}_{f}(s)}{2})^{2}-\tilde{\sigma}%
_{f}(s)+K\sigma_{f}(s),}\tag{17}%
\end{equation}
and
\begin{equation}
\lambda=K+\pi_{f}^{\prime}(s),\tag{18}%
\end{equation}
the $\pi_{f}(s)$ is a polynomial of first degree. The values of $K$ in the
square-root of Eq. (17) is possible to calculate if the expressions under the
square root are square of expressions. This is possible if its discriminate is zero.

\section{Conformable Fractional of the Dependent Temperature Schr\"{o}dinger
Equation}

The SE for two particles interacting via a spherically symmetric (central)
potential $V(r)$ in the N-dimensional space, where $r$ is inter-particle
distance, is given by $\left[  11-13\right]  $%
\begin{equation}
\lbrack\frac{d^{2}}{dr^{2}}+\frac{(N-1)}{r}\frac{d}{dr}-\frac{L(L+N-2)}{r^{2}%
}+2\mu(E-V(r))]\Psi(r)=0,\tag{19}%
\end{equation}
where $L,N,$ and $\mu$ are the angular momentum quantum number, the
dimensionality number and the reduced mass for the quarkonium particle (for
charmonium $\mu=\frac{m_{c}}{2}$ and for bottomonium $\mu=\frac{m_{b}}{2}$),
respectively. Setting the wave function $\Psi(r)=R(r)r^{\frac{1-N}{2}}$, the
following radial SE is obtained
\begin{equation}
\lbrack\frac{d^{2}}{dr^{2}}+2\mu(E-V(r,T)-\frac{(L+\frac{(N-2)}{2})^{2}%
-\frac{1}{4}}{2\mu r^{2}})]R(r)=0.\tag{20}%
\end{equation}
where $V(r,T)$ is the Cornell potential at the finite temperature as in Ref.
$\left[  15\right]  $ and references therein which take following as follows
\begin{equation}
V(r,T)=a\left(  T,r\right)  r-\frac{b\left(  T,r\right)  }{r},\tag{21}%
\end{equation}
where $a\left(  T,r\right)  =\frac{a}{m_{D}(T)r}(1-e^{-m_{D}\left(  T\right)
r})$ and $b\left(  T,r\right)  =be^{-m_{D}\left(  T\right)  r}$ where
$m_{D}\left(  T\right)  $ is the Debye mass that vanishes at $T\rightarrow0$.
$a=0.184$ GeV$^{2}$ and $b$ will determine later. By substituting Eq. (21)
into Eq. (20) and using approximation $e^{-m_{D}\left(  T\right)  r}=%
{\displaystyle\sum\limits_{j=0}^{\infty}}
\frac{\left(  -m_{D}\left(  T\right)  r\right)  ^{j}}{j!}$ up to second-order,
which gives good accuracy when $m_{D}r\ll1$. One obtains%
\begin{equation}
\lbrack\frac{d^{2}}{dr^{2}}+2\mu(E-A+\frac{b}{r}-Cr+Dr^{2}-\frac
{(L+\frac{(N-2)}{2})^{2}-\frac{1}{4}}{2\mu r^{2}})]R(r)=0.\tag{22}%
\end{equation}
where,
\begin{equation}
A=b~m_{D}\left(  T\right)  ,C=a-\frac{1}{2}bm_{D}^{2}\left(  T\right)
,\text{and \ }D=\frac{1}{2}a~m_{D}\left(  T\right)  .\tag{23}%
\end{equation}
By taking $r=\frac{1}{x}$, Eq. (23) takes the following form%

\begin{equation}
\lbrack\frac{d^{2}}{dx^{2}}+\frac{2x}{x^{2}}\frac{d}{dx}+\frac{2\mu}{x^{4}%
}(E-A+bx-\frac{C}{x}+\frac{D}{x^{2}}-\frac{(L+\frac{(N-2)}{2})^{2}-\frac{1}%
{4}}{2\mu}x^{2})]R(x)=0.\tag{24}%
\end{equation}
The expansion of $\frac{C}{x}$ and $\frac{D}{x^{2}}$ in a power series around
the characteristic radius $r_{0}$ of meson up to the second order is given
Ref. $\left[  15\right]  $. The following equation is obtained
\begin{equation}
\lbrack\frac{d^{2}}{dx^{2}}+\frac{2x}{x^{2}}\frac{d}{dx}+\frac{2}{x^{4}%
}(-D_{1}+D_{2}x-D_{3}x^{2})]R(x)=0,\tag{25}%
\end{equation}
where,
\begin{equation}
D_{1}=-\mu(E-A-\frac{3C}{\delta}+\frac{6D}{\delta^{2}}),D_{2}=\mu(\frac
{3C}{\delta^{2}}-\frac{8D}{\delta^{3}}+b),\text{and }D_{3}=\mu(\frac{C}%
{\delta^{3}}-\frac{3D}{\delta^{4}}+\frac{(L+\frac{(N-2)}{2})^{2}-\frac{1}{4}%
}{2\mu}).\tag{26}%
\end{equation}
\ The transition of the conformable fractional of Eq. (25) is obtained as in
Ref. $\left[  23\right]  $%

\begin{equation}
\lbrack D^{\alpha}\left[  D^{\alpha}R(x)\right]  +\frac{2x^{\alpha}%
}{x^{2\alpha}}D^{\alpha}R(x)+\frac{2}{x^{4\alpha}}(-D_{1}+D_{2}x^{\alpha
}-D_{3}x^{2\alpha})]R(x)=0, \tag{27}%
\end{equation}
substituting by Eqs. (7) and (8) into (27), we obtain%

\begin{equation}
R^{\prime\prime}(x)+\frac{\bar{\tau}_{f}(x)}{\sigma_{f}(x)}R^{\prime}%
(x)+\frac{\tilde{\sigma}_{f}(x)}{\sigma_{f}^{2}(x)}R(x)=0, \tag{28}%
\end{equation}
where%

\begin{equation}
\bar{\tau}_{f}(s)=3x^{\alpha}-\alpha x^{\alpha},\sigma_{f}(s)=x^{\alpha
+1},\text{and }\tilde{\sigma}(s)=2(-D_{1}+D_{2}x^{\alpha}-D_{3}x^{2\alpha}).
\tag{29}%
\end{equation}
Hence, the Eq. (28) satisfies Eq. (9). Therefore, Eq. (17) takes the following
form after substituting by Eq. (29)
\begin{equation}
\pi_{f}=-x^{\alpha}+\alpha x^{\alpha}\pm\sqrt{\left(  -x^{\alpha}+\alpha
x^{\alpha}\right)  ^{2}-2(-D_{1}+D_{2}x^{\alpha}-D_{3}x^{2\alpha
})+Kx^{1+\alpha}}. \tag{30}%
\end{equation}
The constant $K$ is chosen such as the function under the square root has a
double zero, i.e. its discriminant equals zero. Hence,
\begin{equation}
K=\left(  \frac{D_{2}^{2}}{2D_{1}}-(1-2\alpha+\alpha^{2}+2D_{3})\right)
x^{\alpha-1}\text{.} \tag{31}%
\end{equation}
Substituting by Eq. (31) into Eq. (30), we obtain%

\begin{equation}
\pi_{f}\left(  x\right)  =-x^{\alpha}+\alpha x^{\alpha}+\frac{D_{2}}%
{\sqrt{2D_{1}}}x^{\alpha}-\sqrt{2D_{1}}\tag{30}%
\end{equation}
The positive sign in Eq. (30) is determined as in Ref. $\left[  15\right]  $.
By using Eq. (13), we obtain%
\begin{equation}
\tau_{f}(x)=3x^{\alpha}+\alpha x^{\alpha}-2\left(  \frac{D_{2}}{\sqrt{2D_{1}}%
}x^{\alpha}-\sqrt{2D_{1}}\right)  ,\tag{24}%
\end{equation}
and using Eq. (14), we obtain
\begin{equation}
\lambda_{n}=\left(  -n\left(  3\alpha-\alpha^{2}\right)  -\frac{2D_{2}\alpha
}{\sqrt{2D_{1}}}-\frac{n\left(  n-1\right)  \alpha\left(  \alpha+1\right)
}{2}\right)  x^{\alpha-1}\text{.}\tag{25}%
\end{equation}
From Eqs. (14 and 18 ); $\lambda=\lambda_{n}$. The energy eigenvalues of Eq.
(22) in the $N$-dimensional space is given%

\begin{align}
E_{nL}^{N}  &  =A+\frac{3C}{\delta}-\frac{6D}{\delta^{2}}-\nonumber\\
&  \frac{2\mu(\frac{3C}{\delta^{2}}+b-\frac{8~D}{\delta^{3}})^{2}}%
{[(2n+1)\pm\sqrt{W+\frac{8\mu C}{\delta^{3}}+4((L+\frac{N-2}{2})^{2}-\frac
{1}{4})-\frac{24\mu D}{\delta^{4}}}]^{2}}. \tag{26}%
\end{align}
with%
\begin{equation}
W=\left(  2n\alpha+\alpha\right)  ^{2}-4\left(  n\left(  3\alpha-\alpha
^{2}\right)  +\frac{1}{2}n\left(  n-1\right)  \alpha\left(  \alpha+1\right)
+\alpha-1\right)  . \tag{27}%
\end{equation}
The radial of wave function takes the following form
\begin{equation}
R_{nL}\left(  r^{\alpha}\right)  =C_{nL}~r^{\left(  -\frac{D_{2}}{\sqrt
{2D_{1}}}-1\right)  \alpha}e^{\sqrt{2D_{1}}r^{\alpha}}(-r^{2\alpha}D^{\alpha
})^{n}(r^{\left(  -2n+\frac{D_{2}}{\sqrt{2D_{1}}}\right)  \alpha}%
e^{-2\sqrt{2D_{1}}r^{\alpha}}). \tag{28}%
\end{equation}
$C_{nL}$ is the normalization constant that is determined \ by $\int\left\vert
R_{nL}\left(  r^{\alpha}\right)  \right\vert ^{2}dr=1$. We note that the
radial wave function in Eq. (28) does not explicitly depend on the number of
dimensions. Hence, $\int\left\vert R_{nL}\left(  r\right)  \right\vert
^{2}dr=1$ remains unchanged.

\section{Discussion of Results}

In this section, the above results are applied on quarkonium masses. the
following relation is used as in Refs. $\left[  12,27\right]  $
\begin{equation}
M=2m+E_{nL}^{N},\tag{29}%
\end{equation}
where $m$ is quarkonium bare mass for the charmonium or bottomonium mesons. By
using Eq. (26), we write Eq. (29) as follows:%
\begin{align}
M_{Q} &  =2m+A+\frac{3C}{\delta}-\frac{6D}{\delta^{2}}-\nonumber\\
&  \frac{2\mu(\frac{3C}{\delta^{2}}+b-\frac{8~D}{\delta^{3}})^{2}}%
{[(2n+1)\pm\sqrt{W+\frac{8\mu C}{\delta^{3}}+4((L+\frac{N-2}{2})^{2}-\frac
{1}{4})-\frac{24\mu D}{\delta^{4}}}]^{2}}.\tag{30}%
\end{align}
Eq. (30) represents quarkonium mass that is calculated in the N-dimensional
space with considering fraction-order, Number of dimensionality, and finite
temperature.. One can obtain the quarkonium masses at zero temperature by
taking $T=0$ leads to $A=D=0$ and $C=a$ and $\alpha=1$. Therefore, Eq. (30)
takes the following form%
\begin{equation}
M_{Q}=2m+\frac{3a}{\delta}-\frac{2\mu(\frac{3a}{\delta^{2}}+b)^{2}}%
{[(2n+1)\pm\sqrt{1+\frac{8\mu a}{\delta^{3}}+4L(L+1)}]^{2}}.\tag{31}%
\end{equation}
Eq. (31) coincides with Ref. $\left[  28\right]  $, in which the authors
obtained the quarkonium mass at zero temperature and $\alpha=1$ in 3D. At
$\alpha=1,$ the following equation is obtained
\begin{equation}
M_{Q}=2m+A+\frac{3C}{\delta}-\frac{6D}{\delta^{2}}-\frac{2\mu(\frac{3C}%
{\delta^{2}}+b-\frac{8~D}{\delta^{3}})^{2}}{[(2n+1)\pm\sqrt{1+\frac{8\mu
C}{\delta^{3}}+4((L+\frac{N-2}{2})^{2}-\frac{1}{4})-\frac{24\mu D}{\delta^{4}%
}}]^{2}}.\tag{32}%
\end{equation}
Eq. (32) is compatible with Ref. $\left[  15\right]  $, in which the author
obtained the quarkonium mass at finite temperature in the N-dimensional space.
To calculate quarkonium mass accordance to Eq. (30), the Debye mass is defined
as in Ref. $\left[  29\right]  $%

\begin{equation}
m_{D}\left(  T\right)  =g(T)T\sqrt{\frac{N_{c}}{3}+\frac{N_{f}}{6}},\tag{34}%
\end{equation}
and%
\begin{equation}
g^{2}(T)=\frac{24\pi^{2}}{\left(  33-2N_{f}\right)  \ln\left(  \frac{2\pi
T}{\Lambda_{MS}}\right)  },\tag{35}%
\end{equation}
with $N_{f}$ and $N_{c}$ as the number of flavours and colors, respectively.
$\Lambda_{MS}=0.1$ GeV and $b=\frac{g^{2}(T)}{3\pi}$. In Figs. (1, 2), the
bottomonium mass is plotted as a function of temperature ratio $\frac{T}%
{T_{c}}$ where $T_{c}$ is the critical temperature equals 170 MeV as in Ref.
$\left[  30\right]  $. In Fig. (1), the bottomonium mass is plotted versus
temperature ratio $\frac{T}{T_{c}}$ in the three-dimensional space. At
$\alpha=1$ which represents the normal case for normal calculus. One note that
the bottomonium mass increases up $T=1.36T_{c}$ then the curve of bottomonium
decreases with increasing temperature. This indicates that the bottomonium is
stable up to $T=1.36T_{c}$ then the bottomonium melts in hot QCD medium. In
Refs. $\left[  6,29,31\right]  ,$ the authors found that the bottomonium melts
above the critical temperature. In Ref. $\left[  6\right]  ,$ the dissociation
of bottomonium $T_{D}\succ$ $Tc$ , $T_{D}=3.2$ $Tc$ \ in Ref. $\left[
29\right]  $, and $T_{D}=1.11$ $Tc$ \ in Ref. $\left[  31\right]  $. In these
works, the SE is solved for quarkonium by using different methods. In
addition, the interpretation is an agreement with Ref. $\left[  27\right]  $,
in which the SE is solved for a nucleon in hot QCD medium. Hence, the present
result is a qualitative agreement with Refs. $\left[  6.29,31\right]  $. To
investigate the effect of fractional parameter, one takes the $\alpha
=0.4,0.6,0.8$. One notes that the curve of bottomonium strongly increases and
then decreases in higher temperatures. In addition, the bottomonium mass is
shifted to higher values by decreasing fractional parameter $\alpha$ which
indicates the binding energy is more bound. Therefore, the dissociation
temperature increases with increasing binding energy. Hence the effect of
fraction parameter supports the bound state of bottomonium. The effect is not
considered in other works.

In this work, one interests to study the effect on the dimensional number on
quarkonium mass. The motivation for this as a natural consequence of the
unification of the two modern theories of quantum mechanics and relativity and
the emergence of the string theory, the investigation of the Standard Model
particles in extra or higher-dimensional space is a hot topic of interest.
From the experimental point of view, the investigation of the existence of
extra dimensions is one of the primary goals of the LHC. The search for extra
dimensions with the ATLAS and CMS detectors is discussed in Ref. $\left[
32\right]  $. \ In Fig. 2, the bottomonium mass is plotted as a function of
temperature ratio for different of fractional parameter at $N=5$. In Fig. 2,
one note that the bottomonium mass decreases with increasing temperature. This
indicates that the bottomonium melts around the critical temperature when the
higher dimensional space is considered and also the quarkonium mass slightly
shifts with increasing fractional parameter. Therefore, the fractional
parameter is a little effect in the higher-dimensional space. In addition, one
notes that the bottomonium mass increases with increasing dimensionality
number. This finding is in agreement with Ref. $\left[  22\right]  $%

\begin{center}
\includegraphics[
height=4.0421in,
width=4.5455in
]%
{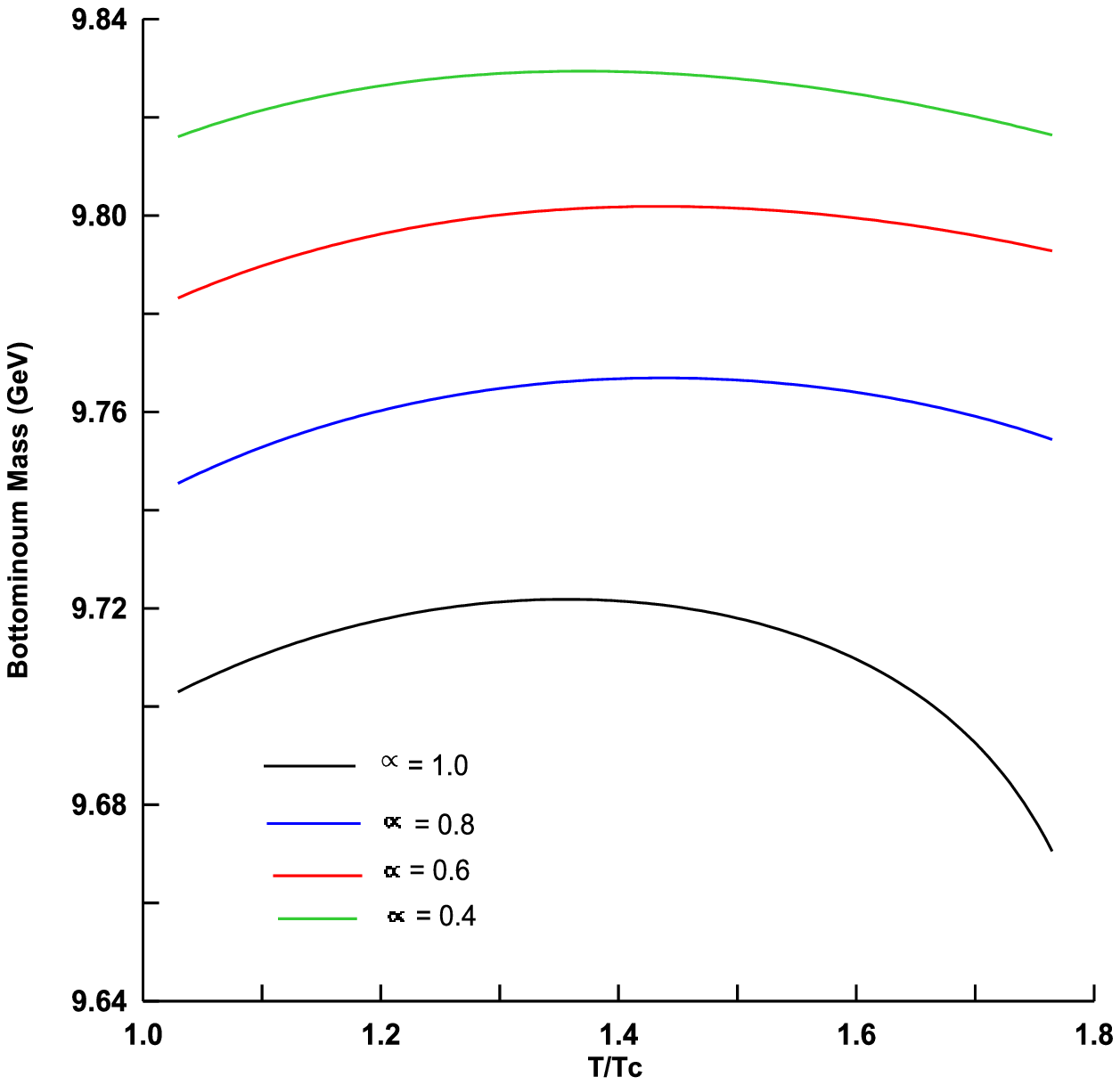}%
\\
\textbf{Fig. 1}: {\protect\small Mass Specrum 1S of bottomonium is plotted as
a function of ratio temperature }$\frac{T}{T_{c}}${\protect\small  \ for
parameters }$m_{b}=4.823${\protect\small \ GeV and }$a=0.184${\protect\small
\ GeV}$^{2}${\protect\small ,at different values of \ fractional parameter at
3D}%
\end{center}
%

\begin{center}
\includegraphics[
height=4.0421in,
width=4.5455in
]%
{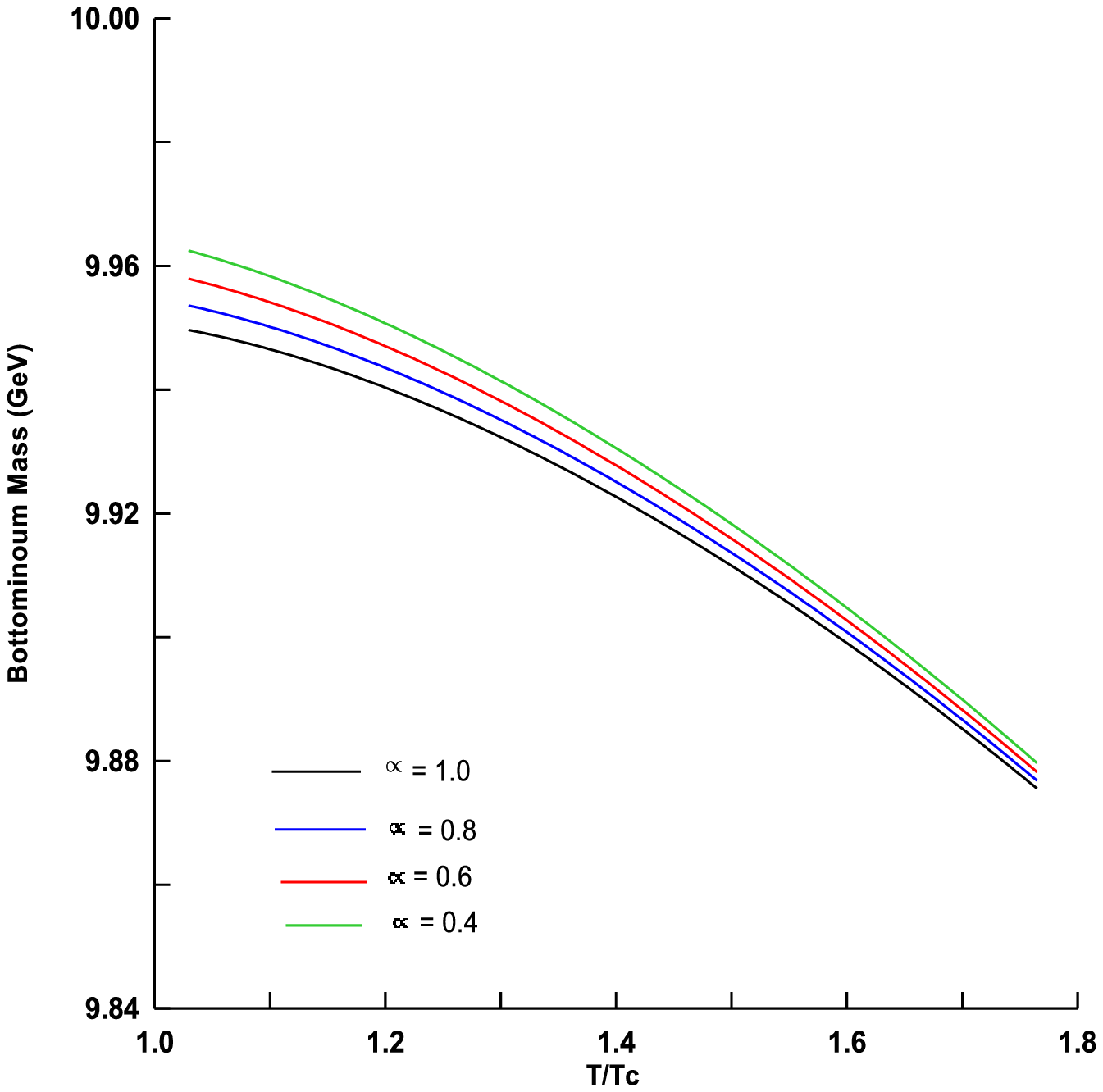}%
\\
\textbf{Fig. 2}: {\protect\small Mass Specrum 1S of bottomonium is plotted as
a function of ratio temperature }$\frac{T}{T_{c}}${\protect\small  \ for
parameters }$m_{b}=4.823${\protect\small \ GeV and }$a=0.184${\protect\small
\ GeV}$^{2}${\protect\small at different values of \ fractional parameter at
5D}%
\end{center}
\ \ \ \ 

In Figs. (3, 4), the charmonium is plotted as a function of temperature ratio
for different values of fractional parameter in 3D and 5D spaces,
respectively. In Fig. (3), one notes that the charmonium mass decreases with
increasing temperature and notes the charmonium melts in given interval of
temperature. This indicates that the charmonium melts under critical
temperature. In Ref. $\left[  31\right]  $, the authors found the 1S state of
charmonuim is melts around 0.99 $T_{c}$. By decreasing fractional parameter,
one notes the charmonium mass increases. In Fig. (4), in 5D, the charmonium
mass decreases with increasing temperature and increases with increasing
dimensionality number. The effect of fractional parameter is a little
sensitive on charmonium when the dimensionality number increases. Thus, the
charmonium mass increases with decreasing fractional parameter and increasing
dimensionality number.%

\begin{center}
\includegraphics[
height=4.0421in,
width=4.5455in
]%
{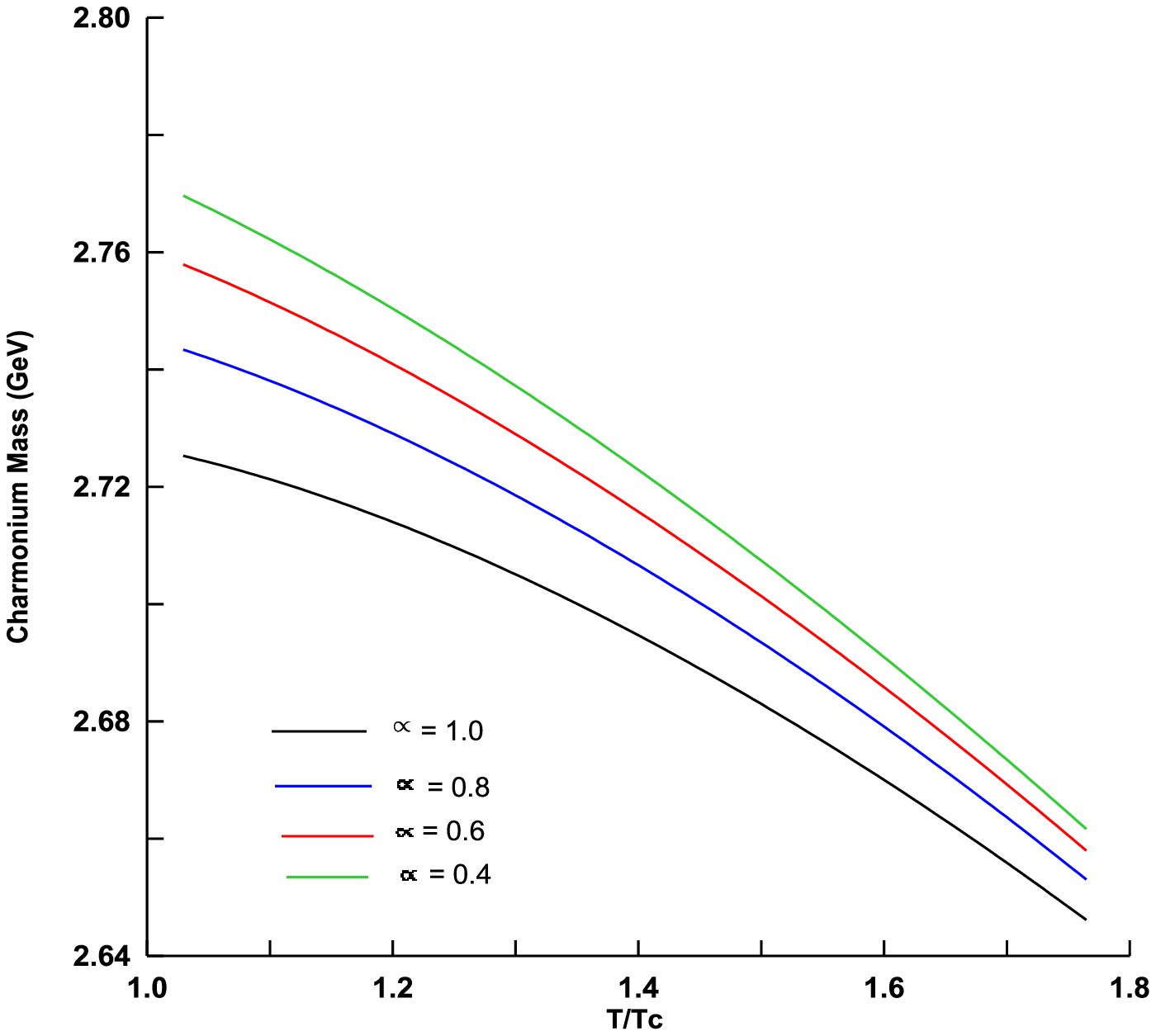}%
\\
\textbf{Fig. 3}: {\protect\small Mass Specrum 1S of charmonium is plotted as a
function of ratio temperature }$\frac{T}{T_{c}}${\protect\small  \ for
parameters }$m_{c}=1.209${\protect\small \ GeV and }$a=0.184${\protect\small
\ GeV}$^{2}${\protect\small at different values of \ fractional parameter at
3D}%
\end{center}
%

\begin{center}
\includegraphics[
height=3.039in,
width=4.5455in
]%
{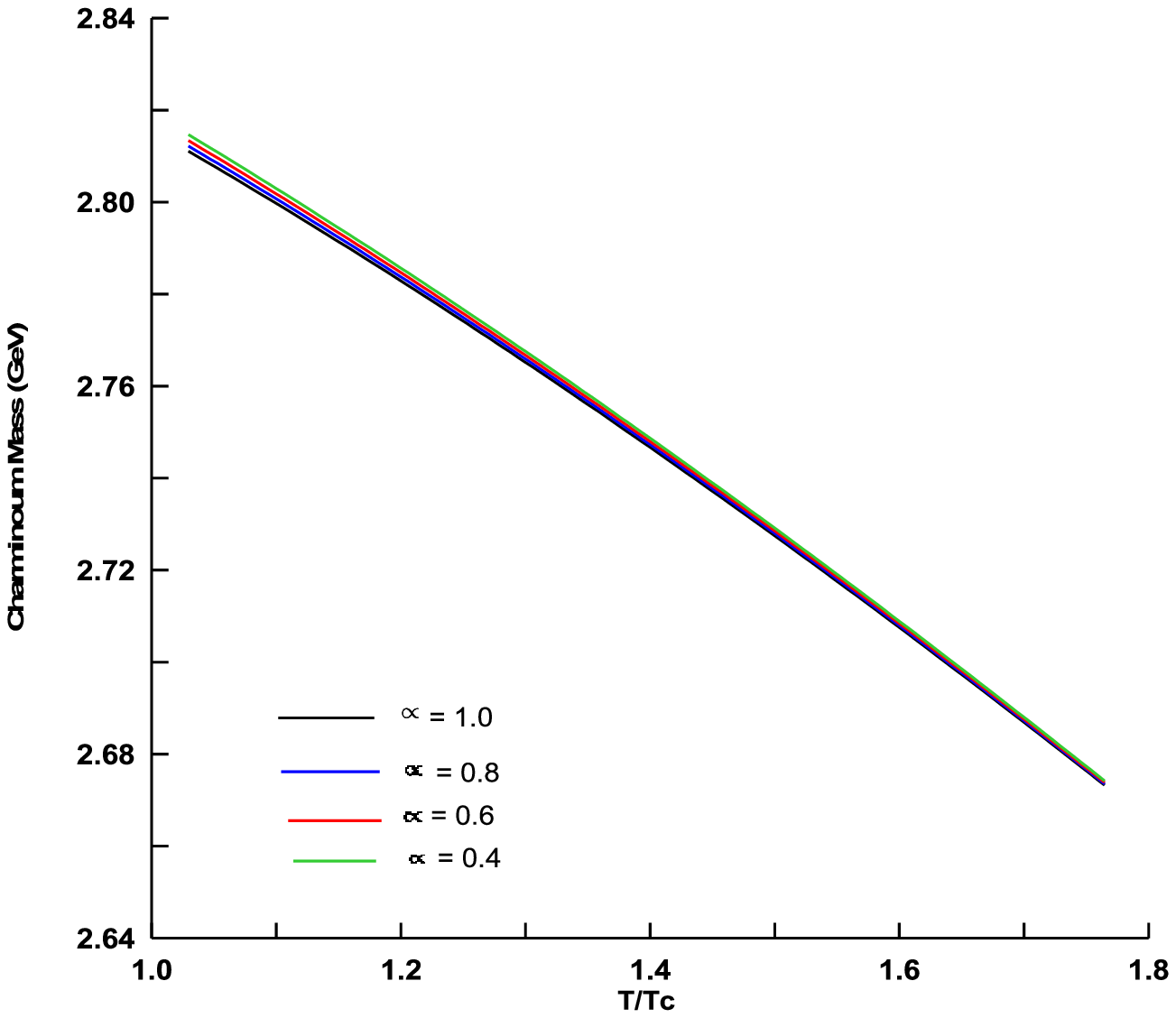}%
\\
\textbf{Fig. 4}: {\protect\small Mass Specrum 1S of charmonium is plotted as a
function of ratio temperature }$\frac{T}{T_{c}}${\protect\small  \ for
parameters }$m_{c}=1.209${\protect\small \ GeV and }$a=0.184${\protect\small
\ GeV}$^{2}${\protect\small  at different values of \ fractional parameter at
5D}%
\end{center}
\ 

\section{Summary and Conclusion}

The CF-NU method is applied to solve N-fractional radial SE. The eigenvalues
of energy and corresponding wave functions are obtained, in which they depend
on the fractional parameter $0<\alpha\leq1$, finite temperature, and
dimensionality number. Particular cases are obtained at $\alpha=1$, $T=0$ and
$T\neq0$ that compatible with recent works. The present results are applied on
the quarkonium system such as charmonium and bottomonium masses in the hot QCD
medium. The present results show that the fractional parameter plays an
important role in the hot QCD medium since the binding energy of charmonium
and bottomonium is more bound by decreasing fractional parameter. In addition,
the effect of fractional parameter is studied in hot medium when the higher
dimensional space is considered $\left(  N=5\right)  $. The results show that
the effect of fractional parameter is a little sensitive on quarkonium masses
at $N=5.$ The dissociation of charmonium and bottomonium is investigated at
$\alpha=1$ which represents a normal case and the agreement is noted in
comparison with other works. One concludes that the fraction calculus plays an
important role to give more information about quark-gluon plasma and expects a
key for solving many theoretical problems in high energy physics.

\begin{enumerate}
\item B. K. Patra and D. K. Srivastava, Phys. Lett. B \textbf{505}, 113 (2001).

\item D. Pal, B. K. Patra, and D. K. Srivastava, Eur. J. Phys. C \textbf{17},
179 (2000).

\item V. Agotiya, V. Chandra,\ and B. K. Patra1, Phys. Rev. C \textbf{80},
025210 (2009).

\item T. Matsui and H. Satz, Phys. Lett. B \textbf{178}, 416 (1986).

\item G. P. Malik, R. K. Jha, and V. S. Varma, Astrophysical J. \textbf{503},
446 (1998).

\item C.-Y. Wong, Phys. Rev. C \textbf{65}, 034902 (2002).

\item F. Arleo, J. Cugnon, and Y. Kalinovsky, Phys. Lett. B \textbf{614}, 44 (2005).

\item C.-Y. Wong, Phys. Rev. C \textbf{76}, 014902 (2007).

\item W. M. Alberico, A. Beraudo, A. D. Pace, and A. Molinari, Phys. Rev. D
\textbf{75}, 074009 (2007).

\item X. Y. Wu, B. Zhang, and X. Liu, Inter. J. Theor. Phys. \textbf{50}, 2546 (2011).

\item M. Abu-Shady and A. N. Ikot, Euro. Phys. J. Plus \textbf{134}, 321, (2019).

\item M. Abu-Shady, T. A. Abdel-Karim, and S. Y. Ezz-Alarab, J. Egyp. Math.
Soci. \textbf{27}, 14, (2019).

\item M. Abu-Shady, T. A. Abdel-Karim, E. M. Khokha, Adv. High Ener.
Phys.\textbf{ 2018}, 7356843, (2018).

\item M. Abu-Shady, E. M. Khokha, Adv. High Ener. Phys. \textbf{2018},
7032041, (2018).

\item M. Abu-Shady, J. Egypt Math. Society \textbf{25}, 86 (2017).

\item E. M. Khokha, M. Abu-Shady, T. A. Abdel-Karim, Inter. J. Theor. App.
Math. \textbf{2}, 86, (2016).

\item M. Abu-Shady, Adv. Math. Phys. \textbf{2016}, 1, (2016).

\item K. B. Oldham and J. Spanier, The Fractional Calculus (Academic Press,
New York, 1974).

\item I. Podlubny, Fractional Differential Equations (Academic Press, 1999).

\item D. Baleanu, K. Diethelm, E. Scalas and J. J. Trujillo, Fractional
Calculus Models and Numerical Methods (World Scientific, 2017).

\item 15. J. L. Wang and H. F. Li, Comput. Math. Appl. \textbf{62}, 1562 (2011).

\item A. Al-Jamel, Int. J. Mod. Phys. A \textbf{34}, 1950054 (2019).

\item H. Karayer, D. Demirhan and F. B\"{u}y\"{u}kk\i l\i\c{c}, Commun. Theor.
Phys. \textbf{66}, 12 (2018).

\item R. Herrmann, arXiv:math-ph/0510099v4

\item M. Caputo and M. Fabrizio, Prog. Fract. Differ. Appl. 1, 73 (2015).

\item R. Khalil, M. A. Horani, A. Yousef and M. Sababheh, J. Comput. Appl.
Math. \textbf{264}, 65 (2014).

\item P. Cheng, Q. Meng, Y. Xia, J. Ping, and H. Zong, Phys. Rev. D
\textbf{98}, 116010 (2018).

\item S. M. Kuchin and N. V. Maksimenko, Univ. J. Phys. Appl. \ \textbf{1},
295 (2013).

\item P. K. Srivastava, O. S. K. Chaturvedi, L. Thakur, Eur. Phys. J. C
\textbf{78}, 440 (2018).

\item M. Y. Jamal, S. Mitra, and V. Chandra, Phys. Rev. D \textbf{97}, 094033 (2018).

\item C.-Y. Wong, J. Phys. Part. Phys. \textbf{28}, 2349 (2002).

\item ATLAS and CMS Collabs. (J. Kretzschmar), Nucl. Part. Phys. Proc. 273275,
541 (2016).
\end{enumerate}

\end{document}